\documentclass[letterpaper,journal]{IEEEtran}
\usepackage{amsmath,amsfonts}
\usepackage{algorithmic}
\usepackage{algorithm}
\usepackage{array}
\usepackage[caption=false,font=normalsize,labelfont=sf,textfont=sf]{subfig}
\usepackage{booktabs,tabularx,array,pifont}
\usepackage{textcomp}
\usepackage{stfloats}
\usepackage{url}
\usepackage{makecell}
\usepackage{verbatim}
\usepackage{graphicx}
\usepackage[percent]{overpic}
\usepackage{cite}
\usepackage{pifont} 
\usepackage{multirow} 
\usepackage{amssymb}  
\usepackage{amsthm}   
\usepackage{ragged2e}
\usepackage[table]{xcolor} 
\newcommand{\rev}[1]{\textcolor{black}{#1}}
\newcommand{\revtag}{\color{black}}
\newcolumntype{Y}{>{\raggedright\arraybackslash}X}
\begin{document}

\title{Adaptive 3D-RoPE: \rev{Channel-Driven} Rotary Positional Embedding for Wireless Foundation Models}

\author{Chenyu Zhang, Xinchen Lyu, Chenshan Ren, Yanzhao Hou, Xuefei Zhang, Shuhan Liu, and Qimei Cui
    \thanks{The work was supported by the National Science and Technology Major Project (2025ZD1302200).}
    \thanks{C. Zhang and X. Zhang are with National Engineering Research Center for Mobile Network
    Technologies, Beijing University of Posts and Telecommunications, Beijing
    100876, China (e-mail: zhangchenyu2024@bupt.edu.cn; zhangxuefei@bupt.edu.cn).}
    \thanks{X. Lyu, Y. Hou, and Q. Cui are with National Engineering Research
    Center for Mobile Network Technologies, Beijing University of Posts
    and Telecommunications, Beijing 100876, China, and also with the Department of Broadband Communication, Pengcheng Laboratory, Shenzhen
    518055, China (e-mail: lvxinchen@bupt.edu.cn; houyanzhao@bupt.edu.cn; cuiqimei@bupt.edu.cn).}
    \thanks{C. Ren is with the Key Laboratory of Ethnic Language Intelligent Analysis and Security Governance of MOE, Minzu University of China, Beijing 100081, China (e-mail: renchenshan06@163.com).}
    \thanks{S. Liu is with China Telecom Corporation Limited Gansu Branch, Gansu 730000, China (e-mail: liush20@chinatelecom.cn).}
    }

\markboth{IEEE Transactions on Wireless Communications}%
{Zhang \MakeLowercase{\textit{et al.}}: Adaptive 3D-RoPE for Wireless Foundation Models}

\maketitle
\begin{abstract}
\rev{Wireless foundation models (WFMs) have emerged as a promising paradigm for unified channel state information (CSI) acquisition across diverse tasks in sixth-generation (6G) networks. Although WFMs significantly outperform task-specific small models, their zero-shot cross-scenario generalization still remains limited for real-world applications. Existing positional embeddings, the sole interface through which self-attention perceives the temporal–frequency–antenna 3D physical coordinates of CSI, fail to capture the highly dynamic and axis-dependent coherence inherent in wireless channels. This paper proposes Adaptive 3D-RoPE, a channel-driven 3D rotary positional embedding framework for WFMs to dynamically align the 3D positional embeddings with the instantaneous coherence state of heterogeneous CSI. The design proceeds in three stages: first, an axis-wise learnable rotary prior independently preserves the temporal, frequency, and antenna coordinate structures; second, a feature-guided rotary modulation module maps the feature-wise standard deviation of visible CSI tokens to compact, sample-adaptive scales; third, identical coordinate offsets induce dynamically adjusted query–key interactions tailored to the instantaneous channel state. Extensive experiments on both simulated and measured datasets validate the effectiveness of Adaptive 3D-RoPE across three complementary dimensions. It reduces NMSE by 10.14, 6.25, and 4.61~dB relative to baselines under antenna, temporal, and frequency scaling, respectively. It transfers effectively to real-world measured CSI and remains robust under imperfect CSI. Finally, it transfers to the independently designed LWM backbone and
beam-prediction task, improving zero-shot Top-1 accuracy by 8.03 percentage points.}
\end{abstract}

\IEEEpeerreviewmaketitle

\section{Introduction}

\begin{figure*}[!t]
\centering
\includegraphics[width=\linewidth]{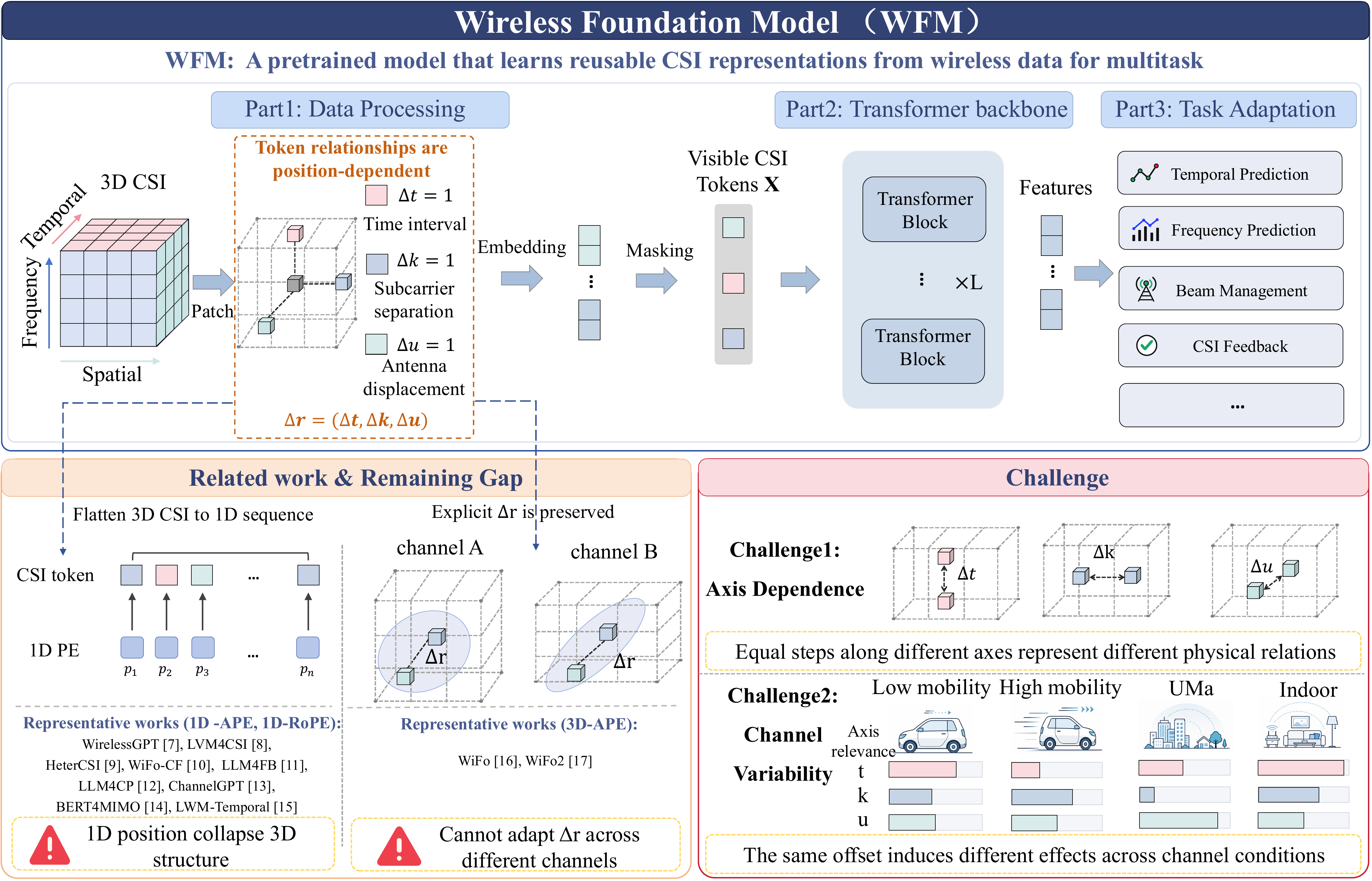}
    \caption{\rev{Overview of the WFM workflow and the motivation for explicit, channel-driven positional embeddings for 3D CSI. The upper part
illustrates CSI tokenization, self-supervised pretraining, and downstream
reuse, while the lower part contrasts existing positional interfaces with
the need to preserve three-axis coordinates and adapt positional interactions
to channel conditions.}}
    \label{fig:pe_comparison}
\end{figure*}

\IEEEPARstart{A}{s} mobile communications advance into the sixth-generation (6G) era, the physical layer is undergoing a profound paradigm shift driven by extreme network densification and high user mobility~\cite{you2021towards}. To sustain reliable connectivity while minimizing signaling overhead, accurately acquiring channel state information (CSI) is paramount. Recently, wireless foundation models have demonstrated significant potential in CSI prediction, i.e., reconstructing full channel matrices from sparse spatial--temporal--frequency observations~\cite{jiang2025csiclip,jiang2026csiclippp,jiang2026csimae}. However, their effectiveness is fundamentally constrained by the exponential dimensionality growth of massive multiple-input multiple-output (MIMO) arrays and ultra-wideband subcarriers, which, coupled with highly non-stationary propagation environments, severely challenges robust cross-scenario generalization~\cite{gao2026ai}.
\subsection{Related Work}
\rev{\textit{1) Wireless Foundation Models:} Wireless foundation models (WFMs)
have emerged as a general-purpose framework for learning channel
representations from large-scale heterogeneous CSI and reusing them across
diverse physical-layer tasks~\cite{alikhani2024lwm,
yang2025wirelessgpt,guo2025lvm4csi,zhang2026hetercsi,liu2025wifocf,
xie2026llm4fb, liu2024llm4cp,yu2024channelgpt,catak2025bert4mimo, alikhani2026lwm,liu2025wifo,liu2025wifo2}. As illustrated in Fig.~\ref{fig:pe_comparison},
a WFM first tokenizes a three-dimensional CSI tensor into patch tokens, where
each token retains its physical coordinate along the temporal, frequency, and
antenna axes. A transformer backbone is then pretrained through self-supervised
objectives, such as masked reconstruction and prediction, to capture reusable
propagation features. The pretrained backbone can subsequently support tasks
including CSI reconstruction, prediction, feedback, and beam management. Unlike
task-specific models optimized for a fixed scenario or dimensionality~\cite{jiang2020deep, jiang2022accurate, xia2019deep,zhou2021informer,cui2025overview}, a WFM must generalize across heterogeneous channel environments. This requirement makes \textit{positional embedding} a fundamental interface through which
self-attention perceives the physical coordinates and captures channel dependence across heterogeneous scenarios.}

\rev{\textit{2) Positional Embeddings:} Because self-attention is permutation-equivariant, positional embedding is required to inject coordinate information into token representations. Absolute positional embedding (APE) adds fixed sinusoidal or learnable position vectors to token embeddings~\cite{vaswani2017attention}. While simple, APE is tied to the observed sequence layout and provides limited translation invariance and length extrapolation.  Relative positional embeddings (RPE) instead parameterize token distances explicitly~\cite{shaw2018self,raffel2020exploring}.
Rotary positional embedding (RoPE) further develops this idea
through a multiplicative phase mechanism: position-dependent rotations of the
query and key vectors induce relative-offset-aware attention scores through
their inner product~\cite{su2024roformer}. Recent studies extend rotary
embeddings to structured multidimensional data, including 2D vision, video, and
3D geometric domains~\cite{heo2024rotary,wei2025videorope,wang2024qwen2vl,
liu2025vrope,li2025hope,ostmeier2024liere,schenck2025learning}, while other
works adapt the rotary spectrum for long-context extrapolation~\cite{barbero2024round,chen2025highfreqhope,
chen2023extending,peng2024yarn,ding2024longrope,sun2023xpos}
or input-dependent conditioning~\cite{veisi2025carope}.}

\rev{Despite these advances, the positional embeddings of current WFMs remain
poorly matched to wireless channels. As summarized in Fig.~\ref{fig:pe_comparison}, most representative WFMs still adopt one-dimensional APE (1D-APE) inherited from language-model backbones~\cite{liu2024llm4cp,yang2025wirelessgpt,yu2024channelgpt,catak2025bert4mimo,guo2025lvm4csi,liu2025wifocf,xie2026llm4fb,
zhang2026hetercsi}; WiFo and WiFo2~\cite{liu2025wifo, liu2025wifo2} use static three-dimensional APE
(3D-APE) to preserve the spatial--temporal--frequency layout; and
LWM-Temporal~\cite{alikhani2026lwm} introduces a one-dimensional RoPE
(1D-RoPE) design. These schemes, however, cannot simultaneously preserve the
three-dimensional channel coordinates and adapt the positional relation to the
channel environment. This gap motivates a channel-driven positional embedding
for WFMs that explicitly preserves 3D coordinates while dynamically adjusting
the induced positional offsets according to the observed CSI.}

\subsection{Challenges and Research Problem}

\rev{However, designing a channel-driven positional embedding for WFMs is non-trivial due to the physical heterogeneity of wireless channels. Specifically, \textit{(1) Axis dependence.} Each CSI token carries temporal, frequency, and antenna coordinates. For adjacent CSI patches with $\Delta\mathbf r$ positional offset, they correspond to different physical separations, namely a time interval, a subcarrier spacing, and an antenna displacement. Flattening these axes into a one-dimensional index mixes non-interchangeable physical relations.
\textit{(2) Channel variability.} The effective relevance of a given offset depends on mobility, delay spread, angular spread, and propagation scenario. Hence, the same offset $\Delta\mathbf r=(\Delta t,\Delta k,\Delta u)$ can correspond to different channel relations in different channel environments, so a static positional mapping cannot provide universally suitable positional offsets for all channel environments.}

\rev{These observations lead to the central research question: \textit{How can a positional embedding preserve explicit 3D relative coordinates while dynamically adapting the induced positional offsets to the observed channel environment, so that the pretrained representation can be reused across unseen scales, configurations, and tasks?}}

\subsection{Contribution and Paper Organization}

To capture the highly dynamic and axis-dependent
coherence inherent in wireless channels, this paper proposes \textbf{Adaptive 3D-RoPE}, a channel-driven positional framework for WFMs. Departing from rigid schedules, it dynamically aligns 3D positional embeddings with the instantaneous coherence state of heterogeneous CSI. The design employs a three-stage inductive process: first, an axis-wise learnable rotary structure independently preserves temporal, frequency, and antenna coordinates; second, feature-guided rotary modulation maps the standard deviation of visible CSI tokens to sample-adaptive scales; and third, identical coordinate offsets induce dynamic query--key interactions tailored to the observed channel. By intrinsically tracking channel variations, this framework establishes a robust foundation for reliable cross-scenario generalization. The key contributions are as follows:
\begin{itemize}
    \item We reveal that identical CSI patch offsets can exhibit different
    empirical CSI dependencies across heterogeneous channel environments,
    highlighting the limitation of static positional mappings. Controlled
    analysis further shows that the learned rotary response follows the same
    trend as empirical CSI dependence, demonstrating adaptive positional
    behavior across channel variations.
    
    \item We formulate the channel-driven positional embedding problem with
    the objective of learning a shared 3D rotary frequency bank with
    sample-adaptive modulation. The formulation preserves the explicit
    temporal, subcarrier, and antenna structures of CSI while adapting
    positional interactions to visible CSI representations.
    
    \item We develop Adaptive 3D-RoPE through three stages: an
    axis-wise learnable rotary frequency bank that preserves temporal,
    subcarrier, and antenna coordinates, feature-guided modulation that maps
    visible-token standard deviations to sample-adaptive scales, and rotary
    attention that applies these scales to query--key relative phases.
\end{itemize}

\rev{
Extensive experiments evaluate Adaptive 3D-RoPE from three complementary perspectives:
\textit{1) Generalization:} It generalizes across unseen CSI scales and scenarios, reducing NMSE relative to 3D-APE by 10.14, 6.25, and 4.61~dB for antenna, temporal, and frequency scaling, respectively.
\textit{2) Practical Resilience:} It maintains its prediction advantage under Sionna-simulated imperfect CSI and supports data-efficient adaptation to real-world measured CSI.
\textit{3) Model-Agnostic:} Its positional interface integrates with independently structured transformer backbones and improves zero-shot 16-beam Top-1 accuracy by 8.03 percentage points in the evaluated downstream task. The source code is available at the GitHub repository~\footnote{https://github.com/zcy8998/adaptive-3d-rope}.}

The remainder of this paper is organized as follows. Section II introduces the wireless-foundation-model setting and positional preliminaries. Section III details the proposed Adaptive 3D-RoPE framework. Section IV presents the experimental results and analysis, followed by conclusions in Section V.

\section{\rev{Preliminaries on Wireless Foundation Models}}

This section first introduces wireless foundation models, including data processing, self-supervised transformer pretraining, and task-specific adaptation. It then reviews how absolute and
rotary positional embeddings incorporate token coordinates into attention,
establishing the positional interface used by the proposed method.

\subsection{\rev{Wireless Foundation Model}}

\rev{A wireless foundation model learns a general channel representation from
large-scale heterogeneous wireless data and reuses it across diverse CSI configurations
and downstream tasks~\cite{jiang2025csiclip,jiang2026csiclippp,jiang2026csimae}.
As illustrated in the upper part of Fig.~\ref{fig:pe_comparison}, a typical WFM
comprises data processing that converts CSI into structured tokens, a
transformer backbone that learns reusable channel features, and task-specific
heads that adapt those features to different wireless tasks.}

\begin{itemize}
    \item \rev{\textbf{Data Processing:}
    Each CSI is represented by a tensor
    $\mathbf{H}\in\mathbb{C}^{T\times K\times U}$, where $T$, $K$, and $U$ denote the temporal, frequency, and antenna dimensions, respectively. CSI may differ in tensor size, carrier configuration, mobility, and propagation scenario. After normalization with training-domain statistics, the real and imaginary parts are concatenated as $\widetilde{\mathbf H}\in\mathbb{R}^{T\times K\times U\times2}$ and partitioned into non-overlapping patches of size $(P_T,P_K,P_U)$. Define the patch-grid sizes as}
    \begin{equation}
    \rev{T_p=\left\lceil\frac{T}{P_T}\right\rceil,\qquad
    K_p=\left\lceil\frac{K}{P_K}\right\rceil,\qquad
    U_p=\left\lceil\frac{U}{P_U}\right\rceil.}
    \revtag
    \label{eq:patch_grid_sizes}
    \end{equation}
    \rev{These patches are linearly projected to form the token sequence $\mathbf{X}=[\mathbf{x}_0,\ldots,\mathbf{x}_{L-1}]^\top\in\mathbb{R}^{L\times D}$, where $D$ is the embedding dimension and $L=T_p K_p U_p$ is the sequence length. Each token $\mathbf{x}_m$ retains the 3D physical coordinate $\mathbf{r}_m=(t_m,k_m,u_m)$ indexing its discrete location within the $T_p \times K_p \times U_p$ patch grid.}

    \item \rev{\textbf{Transformer Backbone:}
    The backbone is typically pretrained via a masked autoencoder (MAE) objective~\cite{he2022masked}, utilizing visible CSI tokens to reconstruct masked regions without task-specific labels. Pretraining on large-scale CSI datasets enables the backbone to capture reusable propagation features, such as local correlations, long-range channel dynamics, and axis-dependent coherence, rather than overfitting to a single scenario. To support this, positional embeddings provide the explicit coordinate structure required by self-attention, ensuring that learned features remain consistent across varying tensor configurations and facilitating zero-shot generalization in unseen channel environments.}

    \item \rev{\textbf{Task Adaptation:}
    A unified pretrained backbone can seamlessly support multiple task-specific heads by extracting generalized channel representations and mapping them to downstream objectives. This architecture enables flexible deployment: in data-scarce scenarios, adaptation requires only fine-tuning a lightweight head atop a frozen backbone, whereas sufficient target data allows for joint end-to-end optimization. By adopting this one-backbone, multi-head paradigm, we eliminate redundant training efforts and allow diverse wireless tasks to benefit from a shared foundation of pretrained channel knowledge.}

\end{itemize}

\subsection{Absolute and Rotary Positional Embeddings}

Self-attention is inherently permutation-equivariant and requires an explicit positional interface to distinguish token coordinates. APE~\cite{vaswani2017attention} addresses this by adding a fixed or learnable position vector $\mathbf{p}_m\in\mathbb{R}^{D}$ to the token embedding $\mathbf{x}_m$ before the query and key projections:
\begin{equation}
\mathbf{q}_m^{\mathrm{APE}} = \mathbf{W}_q (\mathbf{x}_m + \mathbf{p}_m), \quad \mathbf{k}_n^{\mathrm{APE}} = \mathbf{W}_k (\mathbf{x}_n + \mathbf{p}_n).
\label{eq:ape_definition}
\end{equation}
where $\mathbf W_q$ and $\mathbf W_k$ are the learnable query and key projection matrices, respectively.
While straightforward, APE relies on absolute sequence indices, which provides limited length extrapolation capability and fails to explicitly model the relative distance between tokens.

In contrast, standard RoPE~\cite{su2024roformer} injects positional information by rotating the projected query 
($\bar{\mathbf q}_m=\mathbf W_q\mathbf x_m$) 
and key 
($\bar{\mathbf k}_n=\mathbf W_k\mathbf x_n$) 
in paired two-dimensional subspaces. Specifically, the rotated vectors are formulated as 
$\mathbf q_m=\mathbf R_{\boldsymbol\theta,m}\bar{\mathbf q}_m$ 
and 
$\mathbf k_n=\mathbf R_{\boldsymbol\theta,n}\bar{\mathbf k}_n$. 
Here, 
$\mathbf R_{\boldsymbol\theta,m}\in\mathbb R^{d_h\times d_h}$ 
represents the block-diagonal rotation matrix parameterized by a predefined rotary frequency vector 
$\boldsymbol\theta\in\mathbb R^{d_h/2}$, 
with $d_h$ denoting the dimension of each attention head. 
A fundamental property of RoPE is that its query--key inner product reduces to a relative rotation:
\begin{equation}
\mathbf{q}_m^\top \mathbf{k}_n = \bar{\mathbf{q}}_m^\top \mathbf{R}_{\boldsymbol\theta,n-m} \bar{\mathbf{k}}_n.
\label{eq:rope_relative_property}
\end{equation}
This equation demonstrates that the positional interaction depends exclusively on the scalar relative offset $(n-m)$, where the predefined rotary frequencies determine how this distance shifts the rotary phases.

\rev{However, directly applying standard RoPE to wireless foundation models introduces inherent dimensional conflicts. Because standard RoPE operates on a 1D sequence, a 3D CSI token coordinate $\mathbf{r}_m = (t_m, k_m, u_m)$ must be flattened into a scalar index (e.g., $m = t_m K_p U_p + k_m U_p + u_m$). This scalar difference irreversibly mixes the temporal, frequency, and antenna displacements, thereby erasing their distinct physical propagation semantics. Furthermore, even if one extends RoPE to a static 3D formulation to preserve the explicit vector offset $\Delta\mathbf{r}_{mn} = \mathbf{r}_n - \mathbf{r}_m$, the resulting relative phase shifts would remain rigorously fixed by a handcrafted rotary frequency schedule. Because the effective relevance of a specific offset heavily depends on instantaneous channel environments, any static positional mapping will inherently fail to generalize across heterogeneous wireless environments. These fundamental limitations necessitate the development of an adaptive, channel-driven 3D positional embedding.}

\section{\rev{Channel-Driven Rotary Positional Embedding}}

\rev{This section develops the Adaptive 3D-RoPE framework via three modules: a learnable axis-wise rotary structure to capture the 3D geometry of heterogeneous CSI, feature-guided rotary modulation to dynamically scale this structure per sample using visible-token standard deviations, and channel-driven rotary attention to apply the conditioned bank to explicit 3D offsets for tailored query--key rotations.}

\subsection{\rev{Problem Formulation and Design Rationale}}
\label{sec:rotary_modulation_problem}

\rev{Our objective is to establish a unified positional mapping that generalizes to unseen and heterogeneous channel environments without scenario-specific retuning. Let $f_{\vartheta}^{\Phi}$ denote the transformer backbone $f_\vartheta$ equipped with the positional mapping $\Phi$. A masking operator $\mathcal M$ is drawn from a distribution $p(\mathcal M)$ that removes a portion of the information from $\mathbf{H}$, while its complement $\overline{\mathcal M}$ selects the masked patches. The mapping $\Phi$ can generate an adaptive rotary frequency bank $\boldsymbol\Omega$ from the visible CSI $\mathbf{H}$ as}
\begin{equation}
\rev{
\boldsymbol\Omega
=
\Phi\!\left(\mathcal M(\mathbf H)\right),
\qquad
\boldsymbol\Omega
\in
\mathbb R^{3\times N_h\times(d_h/2)},
}
\revtag
\label{eq:sample_conditioned_bank_mapping}
\end{equation}
\rev{where the three dimensions correspond to the temporal, subcarrier, and antenna axes, the $N_h$ attention heads, and the $d_h/2$ two-dimensional rotary pairs within each head, respectively. The channel-driven positional embedding problem for wireless foundation models can be formulated as:}
\begin{equation}
\rev{\begin{aligned}
\mathbf P: 
\min_{\vartheta,\Phi}\quad
\mathbb E_{\substack{
\mathbf H\sim p(\mathbf H)\\
\mathcal M\sim p(\mathcal M)}}
\left[
\left\|
\overline{\mathcal M}(\mathbf H)
-\overline{\mathcal M}\!\left(
f_{\vartheta}^{\Phi}\!\left(\mathcal M(\mathbf H)\right)
\right)
\right\|_F^2
\right].
\end{aligned}}
\revtag
\label{eq:shared_masked_objective}
\end{equation}
\rev{Here, $p(\mathbf H)$ denotes the pretraining CSI distribution and $\|\cdot \|_F^2$ denotes the squared Frobenius norm.}
\begin{figure*}[!t]
\centering
\includegraphics[width=\linewidth,keepaspectratio]{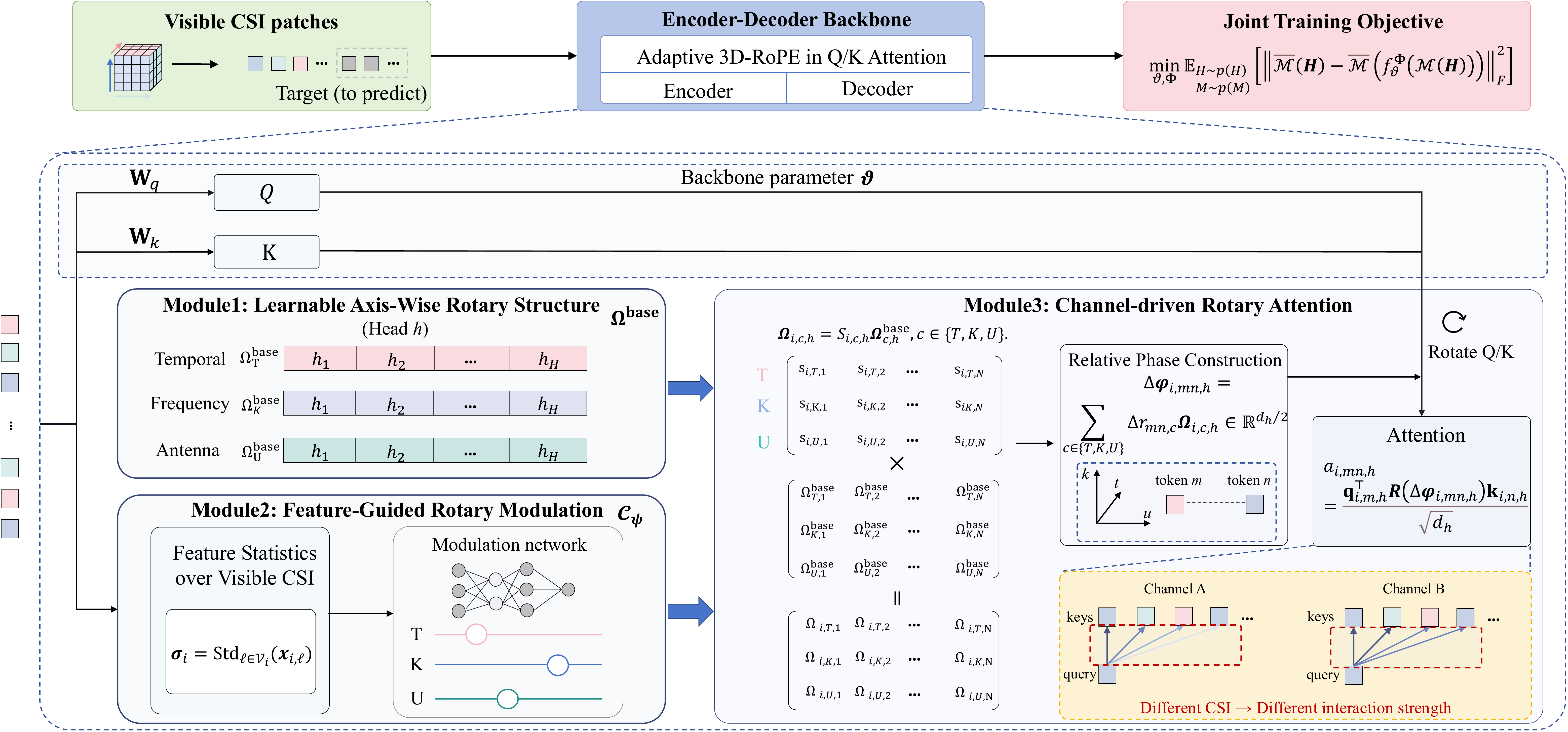}
\caption{\rev{Overview of Adaptive 3D-RoPE. Axis-wise base rotary frequency banks are
jointly learned from masked CSI objectives and modulated by sample-adaptive
axis--head scales derived from the visible-token standard deviation. The
resulting conditioned bank acts on explicit three-dimensional coordinate
offsets to rotate the queries and keys. The encoder and decoder follow the same
construction with independently learned parameters.}}
\label{fig:modulation_schematic}
\end{figure*}

\rev{Problem~\textbf{P} is a coupled non-convex optimization problem since the backbone parameters $\vartheta$ and the positional mapping $\Phi$ jointly determine the reconstruction loss through nonlinear rotary attention. We optimize $\vartheta$ and $\Phi$ jointly through end-to-end self-supervised learning. Beyond optimization, the key design issue is how to parameterize $\Phi$ so that the learned positional mapping simultaneously preserves geometric regularities shared across heterogeneous CSI and adapts to realization-specific channel variations. We therefore separate the positional mapping into a shared axis-wise structure and a realization-specific modulation: a learnable base bank captures the common temporal, frequency, and antenna geometry, while feature-guided modulation derives adaptive scales from visible-token statistics.}

\rev{Fig.~\ref{fig:modulation_schematic} illustrates the resulting architecture and joint training paradigm. This decomposition naturally gives rise to three integrated modules: 1) a learnable axis-wise rotary structure $\boldsymbol\Omega^{\mathrm{base}}$ that captures the shared 3D geometry; 2) feature-guided rotary modulation $\mathcal{C}_\psi$ that adapts the base bank using visible-token statistics; and 3) channel-driven rotary attention that applies the resulting adaptive bank to explicit 3D offsets $\Delta\mathbf{r}=(\Delta t,\Delta k,\Delta u)$. The positional mapping $\Phi$ consists of $\boldsymbol\Omega^{\mathrm{base}}$ and $\mathcal{C}_\psi$: the base bank supplies the shared axis-wise structure, and the network supplies its realization-specific modulation. The encoder and decoder follow the same construction with independent parameters, while each base bank is shared across all transformer blocks within its respective stage.}

\subsection{\rev{Learnable Axis-Wise Rotary Structure}}
\label{sec:axiswise_rotary_structure}

\rev{Standard RoPE employs a predefined rotary frequency vector
$\boldsymbol\theta\in\mathbb R^{d_h/2}$ that determines the phase
variation with positional displacement. Directly extending this
single spectrum to 3D CSI would impose the same prescribed
frequency structure on the temporal, frequency, and antenna axes,
despite their distinct physical semantics and dependency scales.
We therefore generalize the conventional rotary spectrum into an
axis- and head-specific learnable base bank: }
\begin{equation}
\rev{\begin{aligned}
\boldsymbol\Omega^{\mathrm{base}}
&=
\left\{
\boldsymbol\Omega_T^{\mathrm{base}},
\boldsymbol\Omega_K^{\mathrm{base}},
\boldsymbol\Omega_U^{\mathrm{base}}
\right\},\\[-1mm]
\boldsymbol\Omega_c^{\mathrm{base}}
&\in
\mathbb R^{N_h\times(d_h/2)},
\qquad c\in\{T,K,U\}.
\end{aligned}}
\revtag
\label{eq:base_rotary_bank}
\end{equation}

\rev{Compared with the single predefined spectrum in standard
RoPE, this formulation introduces two levels of structural
flexibility. First, the three physical axes maintain independent
rotary spectra, allowing temporal, frequency, and antenna
coordinates to retain their non-interchangeable positional
semantics. Second, each attention head $h$ is assigned its own
frequency vector
$\boldsymbol\Omega_{c,h}^{\mathrm{base}}
\in\mathbb R^{d_h/2}$ along axis $c$, enabling different heads to
capture positional dependencies at different effective scales.
The rotary-pair dimension further provides multiple frequency
components within each axis--head combination.}

\rev{The base banks are directly registered as trainable model
parameters and initialized with the standard RoPE exponential
schedule. Rather than fixing the rotary spectrum a priori, they
are subsequently optimized together with the transformer
backbone and the feature-guided modulation network through end-to-end
self-supervised pretraining. This allows the rotary frequencies of the
three axes and individual attention heads to be learned directly
from heterogeneous CSI, without introducing an auxiliary
frequency-fitting objective or explicit physical supervision.}

\rev{Note that the size of this parameterization, $3N_h(d_h/2)$, is independent of the CSI dimensions ($T$, $K$, $U$), the corresponding patch-grid sizes, and the resulting sequence length $L$. The model therefore learns rotary frequencies rather than a position-indexed embedding table. When the CSI scale changes,
the same learned base bank can be applied directly to newly
encountered coordinates without resizing or extrapolating
positional parameters. This scale-independent shared structure
provides the positional basis for zero-shot generalization.}

\subsection{\rev{Feature-Guided Rotary Modulation}}
\label{sec:feature_guided_modulation}

\rev{While the learned axis-specific rotary structure provides a shared geometric foundation, the relevance of temporal, frequency, and antenna offsets varies across channel environments. We therefore introduce feature-guided rotary modulation, which derives channel-adaptive rotary scales from visible CSI statistics. For the $i$-th CSI sample, let
$\mathbf{X}_i = (\mathbf{x}_{i,\ell})_{\ell=1}^{L_i}$
represent the token sequence obtained after patching and embedding.
Here, $L_i$ indicates the number of patch tokens, and
$\mathcal{V}_i$ denotes the index set of visible tokens. We compute the feature-wise mean and standard deviation over the visible representations as}
\begin{equation}
\rev{\begin{aligned}
\boldsymbol\mu_i
&=
\frac{1}{|\mathcal V_i|}
\sum_{\ell\in\mathcal V_i}
\mathbf x_{i,\ell},\\
\boldsymbol\sigma_i
&=
\left[
\frac{1}{|\mathcal V_i|}
\sum_{\ell\in\mathcal V_i}
\left(
\mathbf x_{i,\ell}-\boldsymbol\mu_i
\right)^{\odot 2}
\right]^{1/2}.
\end{aligned}}
\revtag
\label{eq:visible_token_statistics}
\end{equation}
\rev{Here, $(\cdot)^{\odot 2}$ denotes the Hadamard element-wise square and the square-root operation is applied element-wise. The vector $\boldsymbol\mu_i$ provides the feature-wise centering term, whereas $\boldsymbol\sigma_i$ characterizes the dispersion of the visible representations and serves as the conditioning descriptor for rotary modulation.}

\rev{Before modulation, we standardize the visible-token dispersion
$\boldsymbol\sigma_i$ using fixed, element-wise statistics estimated from the
heterogeneous pretraining data}
\begin{equation}
\rev{
\widetilde{\boldsymbol\sigma}_i
=
\frac{
\boldsymbol\sigma_i-\operatorname{Mean}_{\mathrm{train}}(\boldsymbol\sigma)
}{
\operatorname{Std}_{\mathrm{train}}(\boldsymbol\sigma)
},
}
\revtag
\label{eq:dispersion_normalization}
\end{equation}
\rev{where these fixed training statistics provide a stable and common feature scale across samples, facilitating optimization and ensuring consistent inference.}

\rev{The modulation network is implemented as a two-layer MLP with hidden dimension $d_c$ and GELU activation. The learnable parameters of $\mathcal C_{\psi}$ are
$\mathbf W_1\in\mathbb R^{d_c\times D}$,
$\mathbf b_1\in\mathbb R^{d_c}$,
$\mathbf W_2\in\mathbb R^{3N_h\times d_c}$, and
$\mathbf b_2\in\mathbb R^{3N_h}$.
The $3N_h$ outputs of the second linear layer are organized according to the three physical axes and $N_h$ attention heads. The modulation mapping is given by}
\begin{equation}
\rev{\begin{aligned}
\mathbf z_i
&=
\mathbf W_2
\operatorname{GELU}\!\left(
\mathbf W_1\widetilde{\boldsymbol\sigma}_i
+\mathbf b_1
\right)
+\mathbf b_2,\\[-1mm]
\mathbf S_i
&=
\mathcal C_{\psi}
\left(\widetilde{\boldsymbol\sigma}_i\right)
=
\exp\!\left[
\log(s_{\max})
\tanh(\mathbf z_i)
\right]
\in\mathbb R^{3\times N_h},
\end{aligned}}
\revtag
\label{eq:scale_mapping_compact}
\end{equation}
\rev{where $\mathbf z_i$ is interpreted as a $3\times N_h$ axis--head matrix and
$s_{\max}>1$ is a prescribed maximum modulation factor. The resulting scales satisfy
$s_{\max}^{-1}<S_{i,c,h}<s_{\max}$,
providing bounded positive modulation for each physical axis and attention head. The second linear layer is initialized to zero, yielding $\mathbf z_i=\mathbf 0$ and hence $\mathbf S_i=\mathbf 1$ at initialization, so that the modulation initially recovers the learned base rotary structure.}

\rev{The adaptive rotary frequency bank is then obtained by scaling each axis--head base vector as}
\begin{equation}
\rev{
\boldsymbol\Omega_{i,c,h}
=
S_{i,c,h}
\boldsymbol\Omega_{c,h}^{\mathrm{base}},
\qquad
c\in\{T,K,U\}.
}
\revtag
\label{eq:adaptive_frequency}
\end{equation}
\rev{The encoder and decoder instantiate this construction independently, with separate base banks, normalization statistics, and modulation network parameters. For each encoder or decoder instance, $\mathbf S_i$ and the resulting $\boldsymbol\Omega_i$ are computed once from the visible representations and reused across all attention blocks. Since the modulation statistics are formed exclusively from representations indexed by $\mathcal V_i$, the construction realizes
$\Phi(\mathcal M(\mathbf H))=\boldsymbol\Omega$
without accessing masked CSI.}

\begin{figure*}[!t]
\centering
\includegraphics[width=\textwidth]{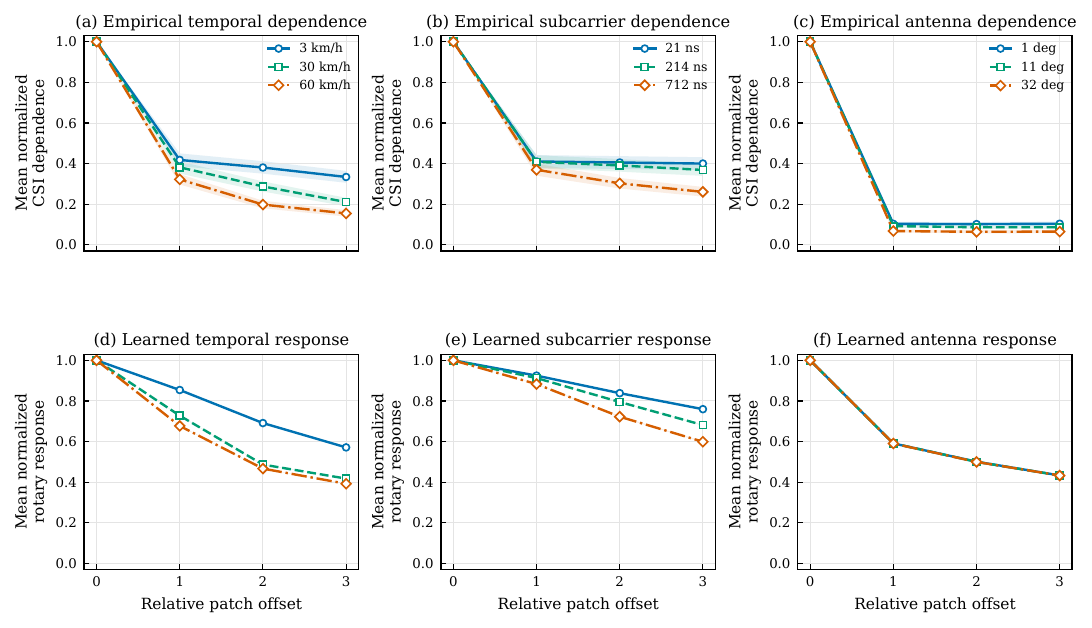}
\caption{\rev{Comparison of empirical CSI dependence and learned rotary response under controlled temporal, subcarrier, and antenna conditions. The two rows use separate normalizations; shaded regions denote 95\% bootstrap confidence intervals.}}
\label{fig:axiswise_dependence_response}
\end{figure*}

\subsection{\rev{Channel-Driven Rotary Attention}}

\rev{Given the sample-adaptive rotary frequency bank $\boldsymbol{\Omega}_{i}$, we apply it to the query--key interaction while explicitly retaining the temporal, frequency, and antenna coordinates. For token $m$ with coordinate $\mathbf r_m=(t_m,k_m,u_m)$, the absolute rotary phase for attention head $h$ is}
\begin{equation}
\rev{\boldsymbol\varphi_{i,m,h}
=\sum_{c\in\{T,K,U\}}
 r_{m,c}\boldsymbol\Omega_{i,c,h}
\in\mathbb R^{d_h/2}.}
\revtag
\label{eq:absolute_phase}
\end{equation}

\rev{For any pair of tokens $m$ and $n$, the content query $\bar{\mathbf q}_{i,m,h}\in\mathbb R^{d_h}$ and key $\bar{\mathbf k}_{i,n,h}\in\mathbb R^{d_h}$ are rotated within paired two-dimensional subspaces:}
\begin{equation}
\rev{\begin{aligned}
\mathbf q_{i,m,h}
&=\mathbf R\!\left(\boldsymbol\varphi_{i,m,h}\right)\bar{\mathbf q}_{i,m,h},\\
\mathbf k_{i,n,h}
&=\mathbf R\!\left(\boldsymbol\varphi_{i,n,h}\right)\bar{\mathbf k}_{i,n,h},
\end{aligned}}
\revtag
\end{equation}
\rev{where $\mathbf R(\cdot)\in\mathbb R^{d_h\times d_h}$ denotes the corresponding block-diagonal rotation matrix.}

\rev{For the explicit 3D coordinate offset $\Delta\mathbf r_{mn}=\mathbf r_n-\mathbf r_m=(\Delta t,\Delta k,\Delta u)$, the induced relative phase is}
\begin{equation}
\rev{\begin{aligned}
\Delta\boldsymbol\varphi_{i,mn,h}
&=\boldsymbol\varphi_{i,n,h}-\boldsymbol\varphi_{i,m,h}\\
&=\sum_{c\in\{T,K,U\}}
\Delta r_{mn,c}\boldsymbol\Omega_{i,c,h}
\in\mathbb R^{d_h/2}.
\end{aligned}}
\revtag
\label{eq:relative_phase}
\end{equation}

\rev{Leveraging the rotation identity $\mathbf R(\boldsymbol\alpha)^\top\mathbf R(\boldsymbol\beta)=\mathbf R(\boldsymbol\beta-\boldsymbol\alpha)$, the scaled attention score elegantly reduces to}
\begin{equation}
\rev{\begin{aligned}
a_{i,mn,h}
&=\frac{\mathbf q_{i,m,h}^{\top}\mathbf k_{i,n,h}}{\sqrt{d_h}}\\
&=\frac{\bar{\mathbf q}_{i,m,h}^{\top}
\mathbf R\!\left(\Delta\boldsymbol\varphi_{i,mn,h}\right)
\bar{\mathbf k}_{i,n,h}}{\sqrt{d_h}},
\end{aligned}}
\revtag
\end{equation}
\rev{where the factor $\sqrt{d_h}$ provides the standard normalization to control the magnitude of the attention logits.}

\rev{This derivation highlights the core intuition of Adaptive 3D-RoPE. As established in Eq. (\ref{eq:relative_phase}), the temporal, subcarrier, and antenna displacements remain mathematically explicit in the phase construction, while their effective contributions are dynamically scaled by the sample-specific bank $\boldsymbol{\Omega}_{i}$. Consequently, identical 3D coordinate offsets can induce distinct relative phase shifts across CSI samples, tailoring the query--key interaction to the observed representation while avoiding the structural collapse of 1D flattening and the rigidity of static positional mappings.}

\subsection{\rev{CSI Dependence and Rotary Response Analysis}}
\label{sec:posthoc_rotary_analysis}

\rev{To examine whether the learned positional dynamics track
axis-dependent variations in CSI dependence,
Fig.~\ref{fig:axiswise_dependence_response} compares two complementary
quantities under controlled temporal, subcarrier, and antenna conditions.
The upper row measures empirical dependence directly from the complex CSI,
whereas the lower row characterizes the positional response induced by the
adaptive rotary frequency bank. For brevity, the sample index is omitted below;
each curve averages the corresponding quantity over 512 test
samples, with 95\% confidence intervals from 10,000 bootstrap resamples. For
$c\in\{T,K,U\}$, let
$\mathbf e_c$ denote the unit vector of axis $c$ and let $\mathbf r=(t,k,u)$
denote a raw CSI-grid coordinate. The empirical dependence at patch-grid offset
$\Delta$ is defined as}

\begin{equation}
\rev{
\widehat C_c(\Delta)
=
\frac{
\left|
\sum_{\mathbf r\in\mathcal R_c(\Delta)}
H(\mathbf r)H^*\!\left(\mathbf r+P_c\Delta\mathbf e_c\right)
\right|
}{
\left[
\begin{aligned}
&\sum_{\mathbf r\in\mathcal R_c(\Delta)}|H(\mathbf r)|^2\\[-1mm]
&\quad\times\sum_{\mathbf r\in\mathcal R_c(\Delta)}
\left|H\!\left(\mathbf r+P_c\Delta\mathbf e_c\right)\right|^2
\end{aligned}
\right]^{1/2}
}.
}
\revtag
\label{eq:empirical_csi_dependence}
\end{equation}
\rev{Here, $P_c$ denotes the patch size along axis $c$, and $(\cdot)^*$ denotes complex conjugation. The set $\mathcal R_c(\Delta)$ contains the raw CSI coordinates
$\mathbf r=(t,k,u)$ satisfying $0\le r_c<N_c-P_c\Delta$ and
$0\le r_{c'}<N_{c'}$ for $c'\ne c$, where
$(N_T,N_K,N_U)=(T,K,U)$. With $P_T=P_K=P_U=4$, we evaluate
$\Delta\in\{0,1,2,3\}$ for each axis and normalize every curve by its
zero-offset value.}

\rev{For each CSI sample, let
$\boldsymbol\Omega_{c,h}\in\mathbb R^{d_h/2}$ denote the adaptive rotary
frequency vector for axis $c$ and attention head $h$. The corresponding
rotary response at patch offset $\Delta$ is defined as}

\begin{equation}
\rev{
G_c(\Delta)
=
\frac{2}{N_h d_h}
\sum_{h=1}^{N_h}
\mathbf 1^\top
\cos\!\left(
\Delta\boldsymbol\Omega_{c,h}
\right),
}
\revtag
\label{eq:rotary_response}
\end{equation}
\rev{where $\mathbf 1\in\mathbb R^{d_h/2}$ and the cosine is applied
element-wise. This response averages the cosine of the relative rotary
phase $\Delta\boldsymbol\Omega_{c,h}$ over all rotary pairs and attention
heads. Eq.~\eqref{eq:rotary_response} is evaluated independently for the
encoder and decoder using their respective rotary banks and attention
dimensions, and the resulting responses are averaged equally in
Fig.~\ref{fig:axiswise_dependence_response}.}

\rev{The paired visualization examines whether variations in empirical
CSI dependence across controlled channel conditions are accompanied by
corresponding changes in the learned rotary response, particularly in
terms of decay direction and condition ordering. The controlled physical
variables are used only to define the evaluation groups and interpret the
observed trends; the proposed method derives its adaptation solely from
the visible CSI representations, without using these variables as explicit
inputs or supervision.}

\rev{Along the temporal axis, increasing mobility generally accelerates
the decay of empirical CSI dependence, while the learned response exhibits
the same overall tendency without implying a strict monotonic relationship
with mobility. Along the subcarrier axis, increasing RMS delay spread leads
to a clear acceleration in the decay of both empirical dependence and the
learned response. Along the antenna axis, both quantities vary only slightly
across the controlled ULA angular-spread conditions. These observations
show that the proposed method preserves the distinct three-axis structure
while adapting its positional response to variations reflected in the
visible CSI.}

\section{Experiments}
This section evaluates whether Adaptive 3D-RoPE can combine efficient positional modeling with robust generalization across heterogeneous wireless channels. We first describe the experimental settings and then examine its generalization, target-domain adaptation, robustness to imperfect CSI, data efficiency, model-agnostic transfer, rotary-modulation ablations, and computational overhead.
\subsection{Experiment Settings}
\label{sec:experiment_settings}

\rev{We conduct our experiments on a workstation equipped with two Intel Xeon Platinum 8468V CPUs, approximately 1.5~TB of RAM, and eight NVIDIA H800 PCIe GPUs. The implementation and execution of model training are carried out using the PyTorch framework~\cite{PaszkeGMLBCKLGA19}.}

\subsubsection{Datasets and System Configurations}
\rev{We use the QuaDRiGa simulator~\cite{jaeckel2014quadriga}
to construct the primary spatial--temporal--frequency CSI datasets. It models a
MISO-OFDM link under the 3GPP TR~38.901 channel model~\cite{3gpp38901},
with a vertically polarized BS uniform planar array using the TR~38.901
element pattern, half-wavelength spacing, and a $7^\circ$
electrical downtilt, serving a single-antenna UE. We further incorporate DeepMIMO v4 O1
~\cite{alkhateeb2019deepmimo}, an outdoor ray-tracing scenario, for
model-agnostic evaluation. Sionna~\cite{hoydis2022sionna} is used to generate
TR~38.901 UMa CSI and simulate sparse-pilot acquisition under imperfect CSI,
and MaMIMO-UAV~\cite{colpaert2023mamimo} provides measured CSI for evaluating real-world transfer. Table~\ref{tab:dataset_summary} summarizes the
data configurations used in the experiments.}

\begin{table*}[t]
\centering
\caption{\rev{Data sources, propagation scenarios, and CSI tensor configurations used for pretraining and evaluation.}}
\label{tab:dataset_summary}
\footnotesize
\setlength{\tabcolsep}{3.5pt}
\renewcommand{\arraystretch}{1.08}
\resizebox{\textwidth}{!}{%
\begin{tabular}{lllcccccc}
\toprule
\rev{Usage} & \rev{Dataset} & \rev{Scenario} & \rev{$f_c$ (GHz)} & \rev{$\Delta f$ (kHz)} & \rev{$\Delta t$ (ms)} & \rev{$T$} & \rev{$K$} & \rev{$U$} \\
\midrule
\rev{Pretraining} & \rev{QuaDRiGa~\cite{jaeckel2014quadriga}} & \rev{Indoor/UMi/UMa/RMa} & \rev{1.5/2.5/4.9/5.9} & \rev{90--360} & \rev{0.5--1} & \rev{16/24} & \rev{32/64/128} & \rev{4/8/16/32} \\
\midrule
\multirow{4}{*}{\rev{Evaluation}}
& \rev{QuaDRiGa~\cite{jaeckel2014quadriga}} & \rev{Indoor/UMi/UMa/RMa} & \rev{0.9/2.1/3.5/6.8/26/28/39/60} & \rev{15--240} & \rev{0.125--1} & \rev{8--64} & \rev{32--1024} & \rev{4--256} \\
& \rev{DeepMIMO v4~\cite{alkhateeb2019deepmimo}} & \rev{O1} & \rev{3.5} & \rev{312.5} & \rev{--} & \rev{1} & \rev{64/128} & \rev{16} \\
& \rev{Sionna~\cite{hoydis2022sionna}} & \rev{UMa} & \rev{3.5} & \rev{30} & \rev{0.5} & \rev{8} & \rev{288} & \rev{4} \\
& \rev{MaMIMO-UAV~\cite{colpaert2023mamimo}} & \rev{UAV measurements} & \rev{2.61} & \rev{180} & \rev{1} & \rev{16} & \rev{100} & \rev{8} \\
\bottomrule
\end{tabular}%
}
\end{table*}

\subsubsection{Architecture and Task Details}
\rev{For the positional-embedding comparison, all variants share the same ViT encoder--decoder backbone and 3D CSI tokenization with $4 \times 4 \times 4$ patches. The encoder comprises eight transformer blocks with a token dimension of 768 and 12 attention heads, while the decoder comprises 4 blocks with a token dimension of 512 and 16 heads; both use an MLP expansion ratio of four. All architectural hyperparameters are kept identical across these variants. In contrast, task-specific baselines retain their original architectures and training protocols. We evaluate the proposed framework on three CSI tasks:
1) CSI reconstruction, 2) time-domain prediction, and
3) frequency-domain prediction. The three tasks can be formulated as}
\begin{align}
\rev{\hat{\mathbf H}}
&\rev{=
f_{\vartheta}^{\Phi}\!\left(
\mathcal M(\mathbf H)
\right),}
\revtag
\\
\rev{\hat{\mathbf H}[T_h+1:T,:,:]}
&\rev{=
f_{\vartheta}^{\Phi}\!\left(
\mathbf H[1:T_h,:,:]
\right),}
\revtag
\\
\rev{\hat{\mathbf H}[:,K_u+1:K,:]}
&\rev{=
f_{\vartheta}^{\Phi}\!\left(
\mathbf H[:,1:K_u,:]
\right).}
\revtag
\end{align}
\rev{where $\mathcal M(\mathbf H)$ denotes the visible CSI under the selected
masking strategy, $T_h$ and $K_u$ are the numbers of observed historical time
slots and input subcarriers, respectively, and $\hat{\mathbf H}$ is the
predicted CSI.}

\subsubsection{Training and Adaptation Protocols}
\rev{\textbf{Pretraining Protocol:} The main model is pretrained on 16
heterogeneous QuaDRiGa configurations, each containing 9,000/1,000/2,000
training/validation/test samples. Random, temporal, and frequency masks use
ratios of 0.85, 0.5, and 0.5, respectively. All positional variants are trained
for 150 epochs with AdamW ($\beta_1=0.9$, $\beta_2=0.95$, weight decay 0.05),
an effective batch size of 128, a learning rate of $5\times10^{-4}$, a
10-epoch warm-up, and cosine decay. The rotary frequency bank and feature-guided
modulation use a 0.1 learning-rate multiplier. The encoder and decoder have
independent banks and modulation modules, each shared across all blocks in its
stage, and the modulation bound is set to $s_{\max}=5$.}

\rev{\textbf{Adaptation Protocol:} Unless otherwise specified below, all evaluations are conducted in the zero-shot setting without target-domain training data. For the target-domain adaptation study in Fig.~5(b), the pretrained encoder--decoder is fine-tuned on the available QuaDRiGa target-domain data for 20 epochs with a learning rate of $10^{-5}$. For the robustness evaluation in Fig.~6, the ideal-history and imperfect-CSI Sionna pipelines are adapted independently for two epochs. For the measured-CSI study in Fig.~7, the pretrained encoder--decoder is fine-tuned on the available MaMIMO-UAV target-domain data for 20 epochs with a learning rate of $10^{-5}$, while CNN, LSTM, and Informer are trained from scratch for up to 100 epochs. For the model-agnostic beam evaluation in Fig.~8, the proposed positional interface is transferred to the LWM backbone. Both LWM and Proposed undergo 50 epochs of reconstruction pretraining at $K=64$, followed by fine-tuning at $K=64$ and zero-shot evaluation at $K=128$.}

\subsubsection{Baseline}

To isolate the effect of positional embedding, all evaluated PE variants are implemented on the shared transformer backbone described above and differ only in their positional injection mechanisms. The specific baselines are detailed as follows:

\begin{itemize}

  \item 1D-APE~\cite{vaswani2017attention}: A fixed sinusoidal absolute positional embedding applied after row-major flattening. It represents the standard one-dimensional positional interface inherited by many language-model-based CSI architectures~\cite{liu2024llm4cp,yang2025wirelessgpt,yu2024channelgpt,catak2025bert4mimo,guo2025lvm4csi,liu2025wifocf,xie2026llm4fb,zhang2026hetercsi}.
  
  \item 3D-APE~\cite{liu2025wifo}: A fixed separable sinusoidal absolute positional embedding over the temporal, frequency, and antenna coordinates. It preserves the explicit CSI layout, but remains an absolute and channel-independent prior.
  
  \item \rev{Fixed 1D-RoPE~\cite{alikhani2026lwm}: The same fixed rotary schedule is applied after row-major flattening, providing a one-dimensional relative-position baseline that does not retain explicit three-axis coordinates.}
  
  \item Fixed 3D-RoPE (extended from~\cite{alikhani2026lwm}): A fixed 3D rotary baseline that assigns predefined base rotary frequencies to the temporal, frequency, and spatial dimensions independently.
  
  \item Learnable 3D-RoPE (extended from~\cite{heo2024rotary}): The base rotary frequencies for the $T$, $K$, and $U$ dimensions are optimized as learnable parameters over the entire training dataset. This baseline serves as the direct predecessor of the proposed adaptive variant.
  
\item \rev{Adaptive 3D-RoPE (Proposed): The shared learnable 3D rotary frequency bank with token-standard-deviation conditioning and axis--head modulation in Eq.~\eqref{eq:adaptive_frequency}.}
\end{itemize}
\rev{Apart from comparisons among positional embedding schemes, we further
evaluate the framework against task-specific baselines in their corresponding
application settings:}
\begin{itemize}
  \item \rev{LSTM~\cite{jiang2020deep}: a two-layer recurrent baseline
  independently trained for the measured-CSI comparison.}
  \item \rev{CNN~\cite{xia2019deep}: a convolutional baseline independently trained for the
  measured-CSI data-efficiency comparison.}
  \item \rev{Informer~\cite{zhou2021informer}: a long-sequence forecasting
  transformer independently trained for the measured-CSI comparison.}
\end{itemize}

\begin{figure*}[!t]
\centering
\includegraphics[width=\linewidth,keepaspectratio]{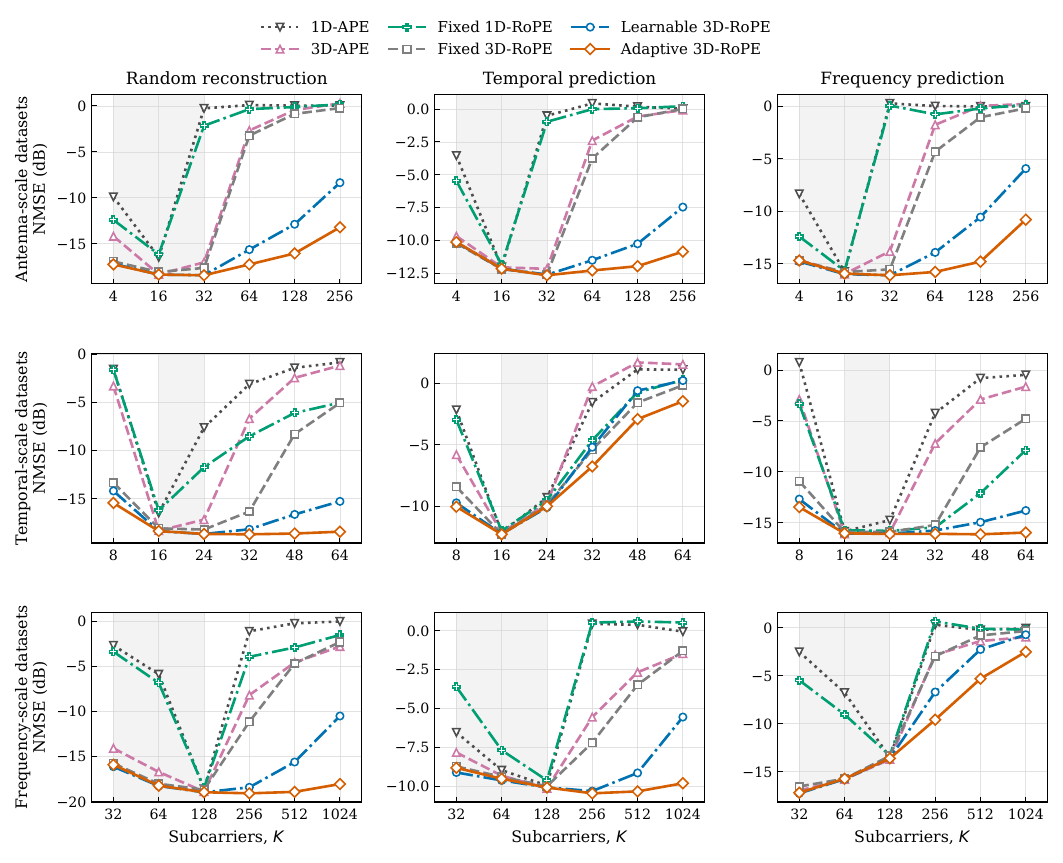}
\caption{\rev{Evaluation of zero-shot generalization across varying antenna, temporal, and frequency scales. The rows denote scaling variations along individual CSI axes, while the columns correspond to the three evaluated tasks: random reconstruction, temporal prediction, and frequency prediction. Gray regions delimit the pretraining scale bounds.}}
\label{fig:extrapolation_generalization}
\end{figure*}

\subsubsection{Evaluation Metric}

Normalized Mean Squared Error (NMSE) is used to quantify the discrepancy between the predicted values and the ground truth~\cite{wen2018deep, xiao2026channel}. It is mathematically defined as:
\begin{equation}
\text{NMSE} = \frac{\| \mathbf{H} - \hat{\mathbf{H}} \|_F^2}{\| \mathbf{H} \|_F^2},
\end{equation}
In our experiments, we report the NMSE in decibels (dB) to better evaluate the reconstruction quality, calculated as:
\begin{equation}
\text{NMSE (dB)} = 10 \log_{10} (\text{NMSE}).
\end{equation}

\subsection{Generalization and Target-Domain Adaptation}
\label{sec:extrapolation_generalization_section}

\rev{\textit{1) Scale Generalization:}
Fig.~\ref{fig:extrapolation_generalization} evaluates whether the positional interfaces learned from the pretraining configurations remain effective when the antenna, temporal, or frequency dimension changes. Aggregated over the three CSI tasks in the linear NMSE domain, Proposed reduces NMSE relative to 3D-APE by 10.14, 6.25, and 4.61~dB on the antenna, temporal, and frequency scales, respectively. Across the three axes and tasks, Proposed remains substantially more stable as the test scale moves beyond the pretraining range, whereas fixed and channel-invariant positional mappings degrade rapidly.}

\rev{At the largest evaluated scale along each axis, we further aggregate performance over the three masked tasks after converting NMSE to the linear domain. The proposed method outperforms Learnable 3D-RoPE by $4.36$, $1.73$, and $2.50$~dB at $U=256$, $T=64$, and $K=1024$, respectively. The proposed positional module plays an important role in preserving accurate spatial–temporal–frequency relationships and effectively mitigating positional mismatch. However, the resulting performance gains vary across tasks, as they are determined not only by the effectiveness of positional adaptation but also by the intrinsic difficulty of each task. When the residual error is increasingly dominated by task-intrinsic factors, such as uncertainty in future CSI evolution or the difficulty of long-horizon forecasting, the scope for further improvement through positional adaptation alone becomes correspondingly limited.}

\begin{figure*}[!t]
\centering
\subfloat[\rev{Zero-shot carrier-frequency generalization.}]{\includegraphics[width=0.485\linewidth]{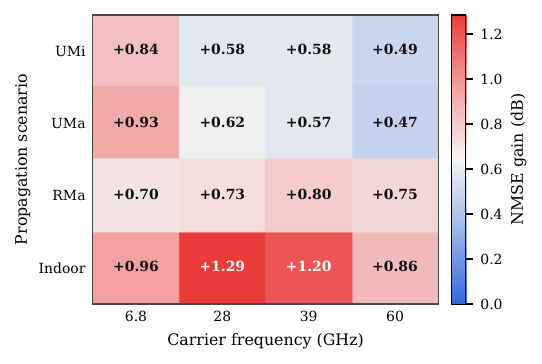}\label{fig:carrier_heatmap}}
\hfill
\subfloat[\rev{Target-domain adaptation at 3.5~GHz.}]{\includegraphics[width=0.485\linewidth]{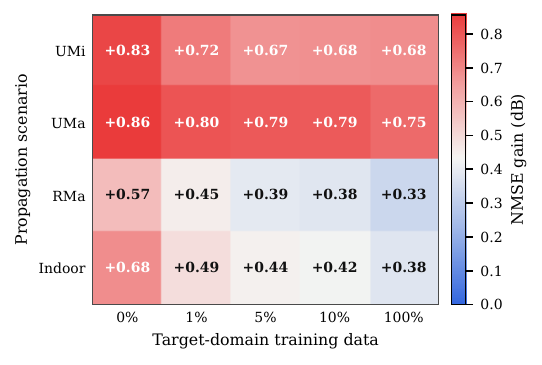}\label{fig:target_adaptation_heatmap}}
\caption{\rev{Comparison of random-reconstruction NMSE gains of Proposed over Fixed 3D-RoPE. (a) Zero-shot carrier-frequency generalization across test carriers and scenarios. (b) Target-domain adaptation at 3.5~GHz. Positive values indicate lower NMSE for Proposed.}}
\label{fig:transfer_heatmaps}
\end{figure*}

\rev{\textit{2) Cross-Band Generalization and Target-Domain Adaptation:}
Fig.~\ref{fig:transfer_heatmaps} shows whether the advantages of adaptive rotary modulation persist under carrier-frequency shifts and subsequent target-domain adaptation. With tensor geometry and mobility held constant, Fig.~\ref{fig:transfer_heatmaps}(a) isolates the zero-shot transfer to unseen carrier frequencies. The Proposed method consistently outperforms Fixed 3D-RoPE across all combinations of carrier frequencies and scenarios, achieving an NMSE gain of up to 1.29~dB. This consistent superiority indicates that the learned adaptive positional rule is not merely memorizing the carrier frequencies observed during pretraining.}

\rev{Fig.~\ref{fig:transfer_heatmaps}(b) illustrates how this advantage evolves as target-domain CSI becomes available at 3.5~GHz. The performance gain is most pronounced in the zero-shot regime and generally narrows as more target data are introduced, since fine-tuning enables both models to better fit the target distribution. This narrowing is more evident in the RMa and Indoor scenarios, whereas the advantage remains relatively stable in UMi and UMa. Note that the proposed method continues to outperform Fixed 3D-RoPE even after complete target-domain adaptation, demonstrating that adaptive rotary modulation is highly beneficial under limited supervision while retaining its effectiveness when abundant target data is available.}

\subsection{Robustness to Simulated Imperfect CSI}

\begin{figure}[!t]
\centering
\includegraphics[width=\linewidth]{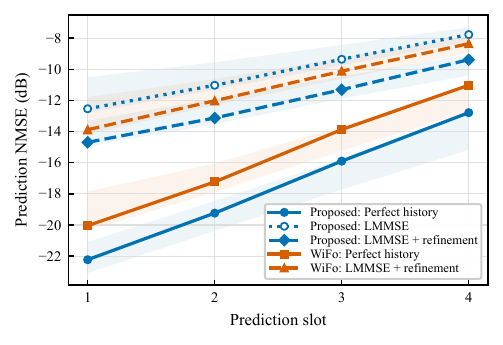}
\caption{\rev{Comparison of temporal CSI prediction from perfect, direct coarse-LMMSE, and refined LMMSE histories in Sionna UMa. Shading indicates 95\% bootstrap confidence intervals.}}
\label{fig:sionna_prediction}
\end{figure}

\rev{In practical communication systems, future CSI prediction is typically performed on imperfect channel estimates rather than ideal ground-truth CSI. To evaluate the impact of channel-acquisition errors on prediction performance, Fig.~\ref{fig:sionna_prediction} considers a Sionna~2.0.1 pipeline using the TR~38.901 UMa model~\cite{hoydis2022sionna}. Each eight-slot CSI window contains four history slots and four prediction targets. Besides prediction from the true past CSI, we evaluate direct prediction from the coarse CSI history obtained through sparse noisy pilots and LMMSE interpolation. The refinement path first refines this same coarse history and then predicts future CSI; both passes use the same adapted backbone. All paths use matched CSI samples, pilot locations, noise realizations, and future prediction targets.}

\rev{Fig.~\ref{fig:sionna_prediction} further examines the propagation and mitigation of acquisition errors in future CSI prediction. Direct prediction from the coarse LMMSE history consistently yields higher NMSE than prediction from the refined history. The shaded 95\% intervals are obtained from 10,000 bootstrap resamples of complete Sionna test groups. With limited adaptation data, the wireless foundation model can reuse its learned CSI representation to refine the imperfect history before prediction, recovering 1.61{}--2.16~dB across the four prediction slots. This denoise-then-predict pipeline reduces acquisition-induced distortions before future CSI prediction without any task-specific architectural modification. The results demonstrate that lightweight adaptation enables the wireless foundation model to alleviate the impact of imperfect CSI and improve the reliability of subsequent prediction. A residual gap to the perfect-history reference remains due to information loss during sparse-pilot acquisition. Proposed also retains lower NMSE than WiFo under both perfect and refined histories, with maximum margins of 2.19{}~dB and 1.19~dB, respectively.}

\begin{figure}[!t]
\centering
\includegraphics[width=\columnwidth]{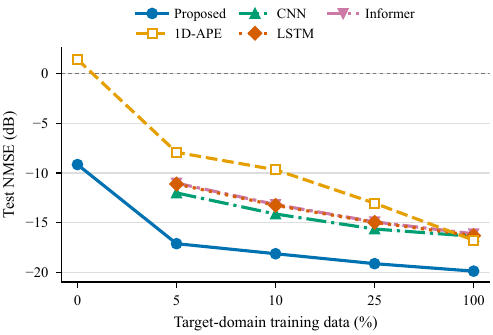}
\caption{\rev{Comparison of measured-CSI data efficiency on MaMIMO-UAV for random masked reconstruction.}}
\label{fig:mamimo_uav_data_efficiency}
\end{figure}

\subsection{\rev{Data Efficiency on Measured CSI}}
\rev{Fig.~\ref{fig:mamimo_uav_data_efficiency} compares the proposed method and 1D-APE with three independently trained small models (CNN, LSTM, and Informer) on real-world measured MaMIMO-UAV CSI. The evaluation encompasses both zero-shot transfer and parallel target-domain fine-tuning across distinct data fractions. The shaded 95\% intervals are obtained from 10,000 bootstrap resamples of measurement parent windows, with the eight array-row crops from each window grouped together; across the displayed points, these intervals span only 0.02--0.24~dB. In the zero-shot regime (0\% target-domain data), only the pretrained models are applicable. The 1D-APE baseline suffers severe degradation with an NMSE of 1.43~dB, whereas the proposed scheme attains -9.15~dB, yielding a substantial 10.58~dB improvement. This contrast highlights the fundamental limitation of a flattened absolute-position embedding when transferring to an unseen measured CSI domain.}

\rev{Under fine-tuning, Proposed and 1D-APE adapt from their pretrained weights, whereas CNN, LSTM, and Informer are trained from scratch on the same masked-reconstruction task. Proposed achieves the lowest NMSE at every data fraction, from -17.12~dB at 5\% data availability to -19.89~dB with full target-domain supervision. Its advantage over the strongest small baseline is 5.11~dB at 5\% and remains 3.49~dB at 25\%. In contrast, 1D-APE remains below all three scratch-trained small models at 5\%, 10\%, and 25\% target data, and surpasses them only under full supervision. Even at 100\%, Proposed retains a 3.11~dB gain over 1D-APE. These results show that the reusable 3D positional representation supports both data-efficient adaptation and continued gains after full target-domain fine-tuning.}

\subsection{\rev{Model-Agnostic Evaluation}}
\label{sec:model_agnostic_evaluation}

\begin{figure}[!t]
\centering
\includegraphics[width=\columnwidth]{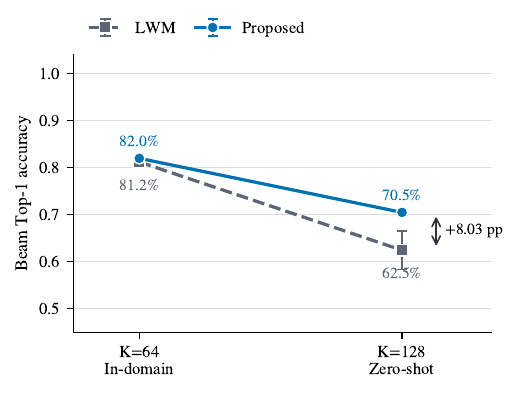}
\caption{\rev{Model-agnostic CSI beam-prediction evaluation on the LWM backbone under the native $K=64$ setting and zero-shot transfer to $K=128$.}}
\label{fig:model_agnostic_evaluation}
\end{figure}

\rev{To evaluate the model-agnostic applicability of the proposed positional interface beyond the primary encoder--decoder backbone, we integrate it into LWM~\cite{alikhani2024lwm}, an independently structured 12-layer encoder-only transformer operating on two-dimensional antenna--frequency patches. While LWM retains its native positional interface, the proposed method replaces it with frequency--antenna rotary banks and feature-guided modulation. For the downstream task, both models employ a shared 16-beam codebook and an identical CSI-to-beam classifier architecture. The complete $K=64$ training split is utilized for adaptation, with the validation split selecting the downstream checkpoint. Furthermore, we evaluate zero-shot beam prediction under a parallel setting where the frequency-token grid is expanded to $K=128$.}

\rev{Fig.~\ref{fig:model_agnostic_evaluation} first shows comparable Top-1 accuracy at the $K=64$ training scale. When the frequency-token grid expands to $K=128$ without further adaptation, Proposed retains 70.50\% accuracy, compared with 62.47\% for LWM, yielding an 8.03{}-percentage-point gain. The error bars denote standard deviations over three independent training runs; their smaller size for Proposed indicates lower run-to-run dispersion in this experiment. These results show that the proposed positional interface can improve downstream beam prediction after integration with a conventional two-dimensional Transformer when the frequency-token grid changes.}

\subsection{\rev{Rotary-Modulation Ablations}}
\label{sec:modulation_ablation_section}

\rev{To isolate the contribution of feature-guided rotary modulation, we conduct controlled ablations in which all variants use the same learnable axis-wise 3D rotary bank and visible-only decoder conditioning, differing only in the presence of modulation and the descriptor used to drive it. The ``None'' variant retains the learnable base bank but applies an identity scale, making it equivalent to Learnable 3D-RoPE without feature-guided modulation. Its comparison with the proposed $\sigma$ variant therefore directly quantifies the gain from sample-adaptive modulation beyond the shared rotary structure. We further compare the visible-token feature mean ($\mu$), standard deviation ($\sigma$), their combinations, and an explicit adjacent-correlation descriptor $\rho$ computed from the visible raw CSI. Specifically, $\rho$ follows the normalized complex-CSI dependence in Eq.~\eqref{eq:empirical_csi_dependence}, but uses a unit raw-CSI lag along each axis and is computed only from adjacent pairs whose two endpoints are visible.}

\begin{table}[!t]
\centering
\caption{\rev{Rotary-modulation input ablation on unseen scales. Each axis-specific score linearly aggregates random reconstruction, temporal prediction, and frequency prediction over the corresponding unseen configurations.}}
\label{tab:modulation_ablation}
\small
\setlength{\tabcolsep}{4.0pt}
\resizebox{\columnwidth}{!}{%
\begin{tabular}{lcccc}
\toprule
\rev{Modulation Input} & \rev{Antenna} & \rev{Time} & \rev{Frequency} & \rev{Overall}\\
\midrule
\rev{None}
& \rev{-11.40}
& \rev{-8.13}
& \rev{-8.30}
& \rev{-9.04}\\

\rev{$\rho$}
& \rev{-12.26}
& \rev{-8.19}
& \rev{-8.53}
& \rev{-9.31}\\

\rev{$\mu$}
& \rev{\textbf{-13.88}}
& \rev{-8.99}
& \rev{\textbf{-10.06}}
& \rev{-10.53}\\

\rev{$\mu+\sigma$}
& \rev{-13.53}
& \rev{-9.46}
& \rev{-10.01}
& \rev{-10.66}\\

\rev{$\rho+\mu+\sigma$}
& \rev{-13.86}
& \rev{-8.85}
& \rev{-10.06}
& \rev{-10.46}\\

\rev{\textbf{$\sigma$ (Proposed)}}
& \rev{-13.74}
& \rev{\textbf{-9.68}}
& \rev{-10.02}
& \rev{\textbf{-10.80}}\\
\bottomrule
\end{tabular}%
}
\end{table}

\rev{Table~\ref{tab:modulation_ablation} summarizes the ablation results over the unseen-scale extrapolation settings in Fig.~\ref{fig:extrapolation_generalization}. For each axis-specific evaluation, the three masked tasks are aggregated in the linear NMSE domain, and the resulting axis-specific scores are equally averaged to obtain Overall. Sample-adaptive modulation consistently improves upon learning the 3D rotary bank alone. In particular, the proposed $\sigma$ variant improves the Overall NMSE by $1.76$~dB over None ($-10.80$ versus $-9.04$~dB), with consistent gains across all three scale evaluations. In contrast, explicitly constructed CSI correlation provides only limited benefit: the $\rho$ variant yields a marginal improvement over None, whereas feature-space statistics lead to substantially larger gains. Although $\mu$ performs slightly better on the Antenna- and Frequency-scale evaluations, $\sigma$ achieves the best Time-scale and Overall performance. Furthermore, augmenting $\sigma$ with additional statistics does not further improve the aggregate result. These results indicate that the visible-token standard deviation provides a compact yet effective descriptor of feature variation for adaptive rotary modulation, whereas $\rho$ is estimated from a smaller set of valid adjacent CSI pairs and is therefore more sensitive to the masking pattern. We therefore use $\sigma$ as the modulation input in the proposed design.}

\subsection{Complexity and Inference Latency}

\rev{We first quantify the incremental computational cost of Adaptive 3D-RoPE. For a token sequence of length $L$ and feature dimension $D$, a standard transformer block has complexity $\mathcal{O}(L^2D+LD^2)$~\cite{vaswani2017attention}, which remains unchanged by the proposed positional interface. The learnable axis-wise rotary bank introduces only $3N_h(d_h/2)$ trainable parameters without additional position-dependent computation. The feature-guided modulation requires a feature-wise standard-deviation scan over the visible tokens followed by two lightweight projections, resulting in an additional complexity of $\mathcal{O}(|\mathcal V_i|D+Dd_c+3N_hd_c)$ for either the encoder or decoder, where $D$ and $N_h$ take the corresponding model dimensions. Since the modulation is computed once and reused across all attention blocks, its additional computation is linear in the number of visible tokens and does not alter the asymptotic complexity of the transformer backbone.}

\rev{Table~\ref{tab:formal_latency} quantifies the practical inference cost under batch-one FP16 inference with $(T,K,U)=(16,64,16)$. On the NVIDIA H800, Adaptive 3D-RoPE increases the median latency by 1.54~ms over Learnable 3D-RoPE, while adding marginal parameters, FLOPs, and peak memory. Relative to learnable 3D-RoPE, the proposed feature-guided modulation incurs only a marginal increase in computational overhead. This slight increment stems primarily from the feature-wise standard deviation calculation across visible tokens and the compact modulation projections, which are executed once and shared across all transformer blocks. Furthermore, the inference latency of the LSTM does not scale linearly with its parameter size, given that its recurrent updates are processed sequentially across temporal slots.}

\rev{These results show that the adaptive positional module adds only a limited incremental cost to the shared Transformer backbone. For the evaluated CSI size, the complete model operates at the millisecond scale on the H800 accelerator. Further deployment studies can therefore examine stricter latency targets by optimizing the overall inference pipeline, including the backbone and runtime system, rather than the positional module alone. Model compression, knowledge distillation, sparse attention, kernel fusion, and specialized inference runtimes are promising directions for this optimization.}

\begin{table}[!t]
\centering
\caption{\rev{Batch-one FP16 inference cost and OOD performance on NVIDIA H800.}}
\label{tab:formal_latency}
\footnotesize
\setlength{\tabcolsep}{3.0pt}
\renewcommand{\arraystretch}{1.04}
\resizebox{\columnwidth}{!}{%
\begin{tabular}{lrrrrrr}
\toprule
\rev{Method}
& \rev{Param.}
& \rev{Mod.}
& \rev{FLOPs}
& \rev{Peak Mem.}
& \rev{Latency}
& \rev{OOD NMSE} \\
& \rev{(M)}
& \rev{(K)}
& \rev{(G)}
& \rev{(MiB)}
& \rev{(ms)}
& \rev{(dB)} \\
\midrule
\rev{1D-APE}
& \rev{69.874}
& \rev{0}
& \rev{22.071}
& \rev{452.5}
& \rev{9.49}
& \rev{-1.98} \\

\rev{3D-APE}
& \rev{69.874}
& \rev{0}
& \rev{22.071}
& \rev{452.5}
& \rev{9.43}
& \rev{-4.00} \\

\rev{Fixed 1D-RoPE}
& \rev{69.874}
& \rev{0}
& \rev{22.072}
& \rev{451.8}
& \rev{11.04}
& \rev{-3.02} \\

\rev{Fixed 3D-RoPE}
& \rev{69.874}
& \rev{0}
& \rev{22.072}
& \rev{451.8}
& \rev{11.08}
& \rev{-5.38} \\

\rev{Learnable 3D-RoPE}
& \rev{69.876}
& \rev{0}
& \rev{22.072}
& \rev{451.8}
& \rev{11.06}
& \rev{-8.95} \\

\rev{Proposed}
& \rev{69.963}
& \rev{87.5}
& \rev{22.073}
& \rev{452.2}
& \rev{12.60}
& \rev{-10.77} \\

\rev{Informer}
& \rev{18.992}
& \rev{0}
& \rev{0.146}
& \rev{162.8}
& \rev{7.45}
& \rev{--} \\

\rev{LSTM}
& \rev{6.302}
& \rev{0}
& \rev{0.189}
& \rev{92.3}
& \rev{10.82}
& \rev{--} \\
\bottomrule
\end{tabular}%
}
\vspace{1pt}

\parbox{\columnwidth}{\footnotesize
\rev{\textit{Note:} Inference cost is measured on the temporal-prediction task with $(T,K,U)=(16,64,16)$ using batch-one FP16 inference. ``Mod.'' reports the parameters of the feature-guided modulation module.}}
\end{table}

%
%

\section{Conclusion}

\rev{This paper presents Adaptive 3D-RoPE as a positional interface for wireless foundation models that preserves explicit temporal--subcarrier--antenna coordinates while adapting query--key interactions from visible CSI. Its shared axis-wise rotary prior and feature-guided rotary modulation are learned jointly from masked CSI objectives, without requiring physical labels or channel-specific side information. Across the simulated antenna-, temporal-, and frequency-scale evaluations, the proposed method reduces NMSE relative to 3D-APE by 10.14, 6.25, and 4.61~dB for antenna, temporal, and frequency scaling, respectively. The gains also persist beyond the main backbone: Proposed improves zero-shot 16-beam prediction at $K=128$ by 8.03 percentage points over LWM and achieves a 10.58-dB zero-shot gain over 1D-APE on measured MaMIMO-UAV CSI. Future work will extend this framework toward neural implicit positional fields, adaptive sampling-aware tokenization, and CSI-native latency optimization that leverages channel sparsity, coherence, and dynamic observation patterns, enabling efficient generalization and real-time inference under irregular, sparse, and high-mobility observations.}




\bibliographystyle{IEEEtran_nodash}
\bibliography{cite}






\end{document}